\documentclass[twocolumn,superscriptaddress,aps,preprintnumbers,amsmath,amssymb,prl,nofootinbib]{revtex4}
\usepackage{bm,latexsym,amsmath,amssymb,amsfonts,mathrsfs}
\usepackage[dvipdfmx]{graphicx}
\usepackage[hang,small,bf]{caption}
\usepackage[subrefformat=parens]{subcaption}
\usepackage{ascmac}
\usepackage{physics}
\usepackage{color,float}
\usepackage{braket}

\newcommand{\simgt}{\lower.5ex\hbox{$\; \buildrel > \over \sim \;$}}
\newcommand{\simlt}{\lower.5ex\hbox{$\; \buildrel < \over \sim \;$}}

\begin{document}

\title{
Feasible generation of gravity-induced entanglement 
by using optomechanical systems
}

\author{Daisuke Miki}
 \email{miki.daisuke@phys.kyushu-u.ac.jp}
\affiliation{Department of Physics, Kyushu University, 744 Motooka, Nishi-Ku, Fukuoka 819-0395, Japan}

\author{Akira Matsumura}
 \email{matsumura.akira@phys.kyushu-u.ac.jp}
\affiliation{Department of Physics, Kyushu University, 744 Motooka, Nishi-Ku, Fukuoka 819-0395, Japan}

\author{Kazuhiro Yamamoto}
 \email{yamamoto@phys.kyushu-u.ac.jp}
\affiliation{Department of Physics, Kyushu University, 744 Motooka, Nishi-Ku, Fukuoka 819-0395, Japan}

\date{\today}
\begin{abstract}
We report the feasibility of detecting the gravity-induced entanglement (GIE)\footnote{GIE can be called quantum gravity-induced entanglement of masses (QGEM) \cite{Thomas,Tilly,Schut}.} with optomechanical systems, which is the first investigation that clarifies the feasible experimental parameters to achieve a signal-to-noise ratio of S/N$=1$. 
Our proposal focuses on GIE generation between optomechanical mirrors,
coupled via gravitational interactions, under continuous measurement, feedback control, and Kalman filtering process, which matured in connection with the field of gravitational wave observations.
We solved the Riccati equation to evaluate the time evolution of the conditional covariance matrix for optomechanical mirrors that estimated the minimum variance of the motions. 
The results demonstrate that GIE is generated faster than a well-known time scale without optomechanical coupling. The fast generation of entanglement is associated with quantum-state squeezing by the Kalman filtering process, which is an advantage of using optomechanical systems to experimentally detect GIE.
\end{abstract}
\maketitle

The verification of gravity-induced entanglement (GIE) is one of the most important milestones in quantum gravity \cite{Bose,Marletto,Krisnanda23,Howl}. Studies by Bose et al. and Vedral and Marletto have stimulated research on
this topic~\cite{Bose,Marletto}, which is regarded as the present-day feasibility study of thought experiments by Feynman (Ref.~\cite{Aspelmeyer2022} and references therein). The authors of Refs.~\cite{Bose,Marletto} proposed an experiment that assumed two particles, each in a spatially localized superposition state, coupled via gravitational interactions.
Such superpositions of massive particles can be generated through the Stern-Gelrach experiment; however, the feasibility of GIE detection using this approach is challenging.
 (e.g., \cite{Wood}). 

Another possible experimental approach for verifying GIE is the use of optomechanical systems in which two mirrors under quantum control are coupled to each other via gravitational interaction \cite{Datta21,Miao20}. 
Optomechanics is a promising field for exploring macroscopic quantum systems \cite{Yanbei,Aspelmeyer,Bowen}.  
Recent experiments have demonstrated that quantum control can realize massive mirrors in the quantum state \cite{matsumoto20,MY,miki23}.
In the previous study \cite{MMY24}, we discussed the conditions for generating GIE using optomechanical systems in the steady-state limit.
We also demonstrated that interactions, except for gravity, were negligible in optomechanical systems.
However, it has not been clarified what kind of experiment is optimal to detect the generation of the GIE. In the current study, we considered this point to determine the optimized setup of an experiment by investigating the signal-to-noise ratio. 
The time evolution of the GIE, which is important for evaluating the feasibility of the experiment, has not been clarified. 
For the first time, we investigated the time evolution of GIE in optomechanical systems under continuous measurements and feedback control as well as with a quantum filtering process.
Under the quantum Kalman filtering process, the time evolution of the entanglement is computed from the solution of the Riccati equation for conditional covariance.
We show that GIE can be generated by reducing thermal noise and realizing a quantum state through the Kalman filtering process~\cite{Yamamoto,Wieczorek}.
The results demonstrated that the generation time was significantly shorter than the well-known time required to generate a GIE without optomechanical coupling \cite{Krisnanda}. 

\begin{figure}[b]
    \centering
    \includegraphics[width=8.7cm]{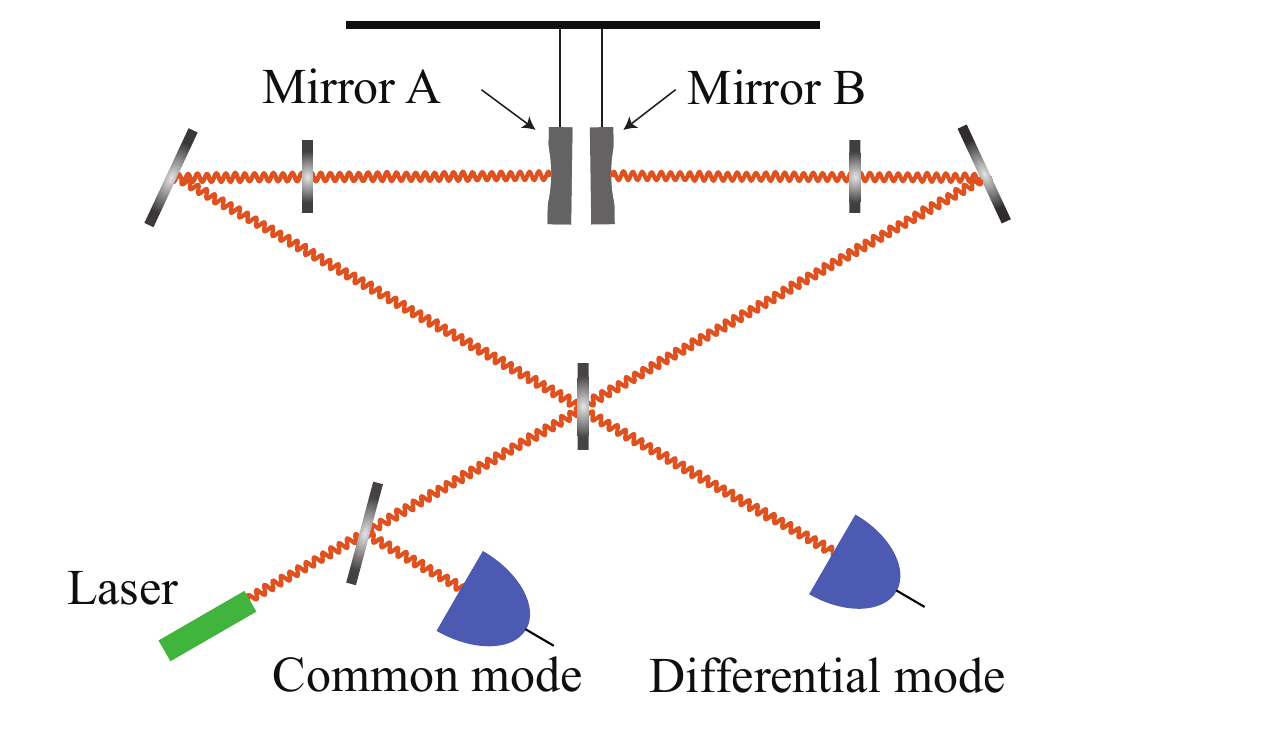}
    \caption{Configuration of the system. 
    We consider two mirrors close to each other to be gravitationally interacting, which are also coupled to cavity lights. The system is constructed with the half mirrors and the reflectors.}
    \label{config2}
\end{figure}

We consider optomechanical systems with two gravitational mechanical mirrors (Fig.~\ref{config2}) whose Hamiltonian is
\begin{align}
    &H=
    \frac{\hbar\Omega}{4}(q_A^2+p_A^2)+
    \frac{\hbar\Omega}{4}(q_B^2+p_B^2)
    +\hbar\omega_ca_A^\dagger a_A
 +\hbar\omega_ca_B^\dagger a_B
 \nonumber\\
    &   
 +\hbar gq_Aa_A^\dagger a_A
    -\hbar gq_Ba_B^\dagger a_B
    -\frac{Gm^2}{L^3}(q_A-q_B)^2,
\end{align}
where $q_j$ and $p_j$ are the mechanical position and momentum operators satisfying $[q_j,p_k]=2i\delta_{jk}$, $a_j$ and $a_j^\dagger$ are the annihilation and creation operator satisfying $[a_j,a_k^\dagger]=\delta_{jk}$, for $j,k=A,B$, 
$\Omega$ is the mechanical frequency, $\omega_c$ is the optical resonance 
frequency in cavities of length $\ell$,
$g=(\omega_c/\ell)\sqrt{\hbar/2m\Omega}$ is the optomechanical coupling, $G$ is the gravitational constant, $m$ is the mass of the mechanical mirrors, and $L$ is the separation between the two mirrors.
Here, we consider the mechanical common mode ($+$) and differential mode ($-$) based on the measurement results and introduce each mode as
    $q_\pm=
    \bigl(q_A\pm q_B\bigr)\sqrt{{\Omega_\pm}/({2\Omega})},~
    p_\pm=
    \bigl(p_A\pm p_B\bigr)    \sqrt{{\Omega}/({2\Omega_\pm})},~
    a_\pm=
    ({a_A\mp a_B})/{\sqrt{2}}$,
where $\Omega_\pm$ is the mechanical frequency of the common and differential modes $\Omega_+=\Omega$ and $\Omega_-=\Omega\sqrt{1-\epsilon}$, respectively. Gravitational coupling is described as 
parameter $\epsilon={4Gm}/{(L^3\Omega^2)}$.

Here, we introduce the optical amplitude quadrature $x_{\pm}=a+a^\dagger$ and optical phase quadrature $y_\pm=(a\pm a^\dagger)/i$ satisfying $[x_\pm,y_\pm]=2i$.
The amplitude and phase quadratures are given by $\sqrt{\kappa}x_\pm=-x_\pm^{\rm in}$ and $\sqrt{\kappa}y=-y_\pm^{\rm in}-4g_m^\pm q_\pm$ for the bad cavity regime $\kappa\gg\Omega$, where $\kappa$ is the optical decay rate, and $x_\pm^{\rm in}$ and $y_\pm^{\rm in}$ are the vacuum noise input \cite{Bowen}.
The motion of the mirrors can be estimated based on phase quadrature measurements.
Using the vector of canonical operators $\bm{r}_\pm=(\hat{q}_\pm,\hat{p}_\pm)^{\rm T}$,
we derive the Langevin equation in matrix form:
\begin{equation}
\dot{\bm{r}}_\pm =\bm{A}_\pm\bm{r}_\pm
+(0,w_\pm)^{\rm T}.
\end{equation}
In the first term, 
$\bm{A}_\pm$ denote 
$2\times 2$ matrix with components 
$(\bm{A}_\pm)_{11}=0$, 
$(\bm{A}_\pm)_{12}=\Omega_\pm=-(\bm{A}_\pm)_{21}$ and 
$(\bm{A}_\pm)_{22}=-\gamma_m$, where $\gamma_m$ denotes the effective mechanical decay rate under feedback.  
The second term with 
$ w_\pm=\sqrt{2\gamma_{m}}p_{\rm in}^\pm -(4g_\pm/\sqrt{\kappa})x_{\text{in}}^\pm$
describes the force noise, where  $p_{\rm in}^\pm$ is the mechanical noise input and $g_\pm=g\sqrt{4P_{\rm in} \Omega /(\hbar\omega_c\kappa \Omega_\pm)}$ is the effective optomechanical coupling. 
This coupling depends on the optical decay rate $\kappa$
and input laser power 
$P_{\rm in}$. 
For evaluating 
$g_\pm$, each optical cavity mode is assumed to have the same average amplitude. 
Optical output quadrature, $Y_\pm$,
is  derived from the optical input-output relation: 
$Y_\pm =\bm{C}_{\pm}\bm{r}_\pm-y^{\rm in}_{\pm}$
, where 
$\bm{C}_\pm= (4g_\pm/\sqrt{\kappa},0)$.

To reduce thermal noise, we employed a quantum filter for the optimal estimation of mechanical motion.
The quantum Kalman filter minimizes the mean squared error
between the canonical operators $\bm{r}_\pm$ and the estimated values $\tilde{\bm{r}}_\pm=(\tilde{q}_\pm,\tilde{p}_\pm)^{\text{T}}$ based on the measurements.
Thus, each component of the covariance matrix conditioned on the measurement results $\bm{V}_\pm=\braket{\{\bm{r}_\pm-\tilde{\bm{r}}_\pm,(\bm{r}_\pm-\tilde{\bm{r}}_\pm)^{\text{T}}\}}$ is minimized \cite{Yamamoto,Wieczorek}.
In the absence of a quantum filter, we only have the average behavior of $\bm{r}_\pm$, and entanglement does not occur because of thermal noise.
Quantum filters are essential for 
reducing thermal fluctuations and 
generate entanglement.
Using quantum Kalman filtering, the conditional covariance matrix follows the Riccati equation
\begin{eqnarray}
\dot{\bm{V}}_\pm=
\bm{A}_\pm\bm{V}_\pm
+\bm{V}_\pm\bm{A}_\pm^{\text{T}}+\bm{N}_\pm-\bm{V}_\pm\bm{C}_\pm^{\text{T}}\bm{C}_\pm\bm{V}_\pm,
\label{eq:Riccati}
\end{eqnarray}
where
$\bm{N}_\pm=\text{diag}[0,n_\pm]$, where the noise term $\bar{n}_\pm=
2\gamma_m(2n_{\text{th}}^\pm+1)+{ 16 g _\pm^2}/{\kappa}$, 
$n_{\rm th}^\pm=k_BT\Gamma/(\gamma_m\hbar\Omega_\pm)$ is the effective thermal phonon number under the feedback control, where 
$k_B$ is the Boltzmann constant, $T$ is the environmental temperature, and $\Gamma$ is the bare mechanical dissipation rate.
Here, we assume that the thermal photon noise is negligible.
The last term describes the effects of the Kalman filtering process, which minimizes all the components of the covariance matrix.
The time required to reach a static solution is approximately proportional to $\kappa/g_\pm^2$.

\begin{figure}
    \centering
    \includegraphics[width=6.7cm]{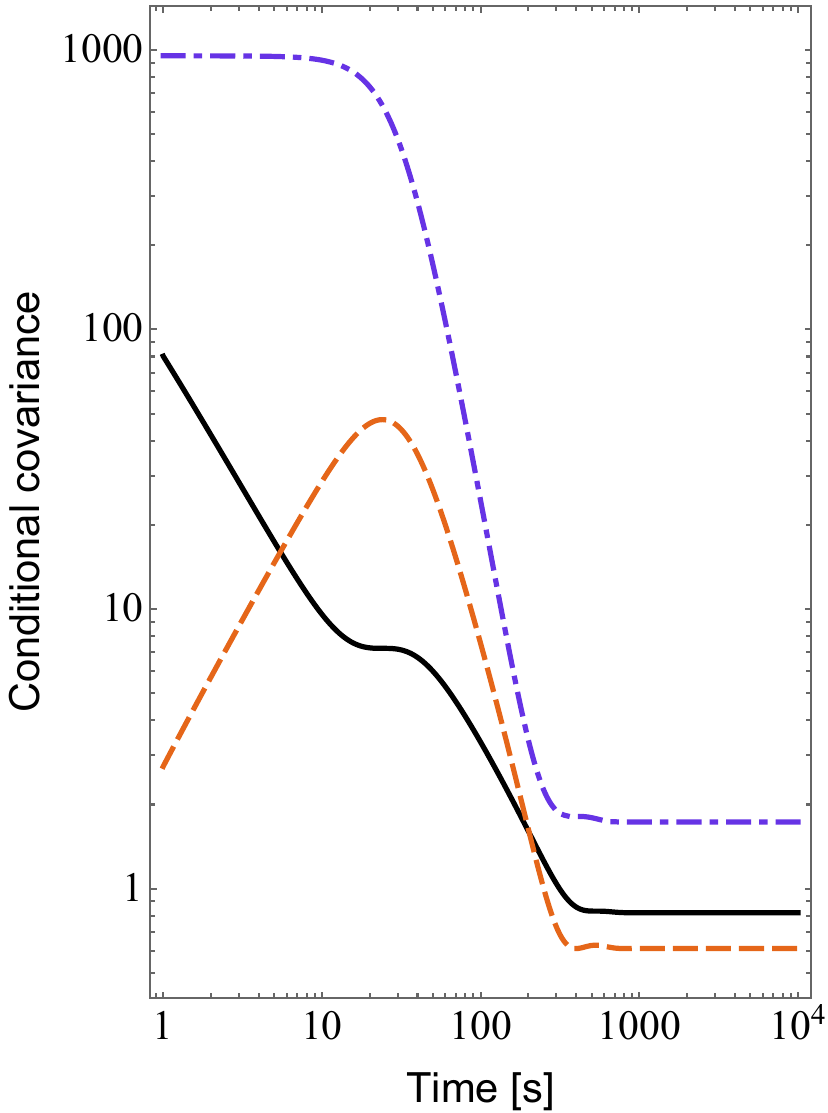}
    \caption{
    Time-evolution of the components of the conditional covariance matrix elements, $V^+_{qq}(t)$ (black solid curve), 
    $V^+_{qp}(t)$ (orange dashed curve), 
    $V^+_{pp}(t)$ (purple dashed-dotted curve).
    The parameters are listed in Table I.
    The behavior of the differential mode is nearly the same. 
    \label{fig:v}}
\end{figure}
\vspace{0mm}
\begin{table}[t]
    \centering
    \begin{tabular}{c|c}
    \hline
~~~~$\Gamma/2\pi$~~~~ & $10^{-18}~{\rm Hz}$
\\ \hline
$\Omega/2\pi$ & $10^{-3}~{\rm Hz}$
\\ \hline
$\gamma_m/2\pi$& $10^{-6}~{\rm Hz}$
\\ \hline
$\kappa/2\pi$ & $10^{8}~{\rm Hz}$
\\ \hline
$\omega_c/2\pi$ & $2.8\times10^{14}~{\rm Hz}$
\\ \hline
$P_{\rm in}$& $0.1~{\rm mW}$
\\ \hline
$\ell$ & $1~{\rm m}$
\\ \hline
$T$ & $1~{\rm K}$
\\ \hline
$\rho$ & $ ~~~~20~{\rm g}/{\rm cm}^3$~~~~
\\ \hline
$m$ & $10^{-1}~{\rm kg}$
\\ \hline
$\Lambda$ & $2$
\\ 
\hline
    \end{tabular}
    \caption{
    Parameters of the numerical solution in Fig.~\ref{fig:v}.
    Ref.~\cite{Belenchia} discussed the experimental situation to achieve low mechanical dissipation.}
    \label{tab:my_label}
\end{table}

We solved the Riccati equation using a numerical method, assuming that the two mirrors were initially in a separable thermal state:
$\bm{\mathcal{V}}(0)=(\bar{n}_+/2\gamma_m)\bm{1}_4$
(see Eq.~\eqref{bmV} for $\bm{\mathcal V}$).
Fig.~\ref{fig:v} shows the components of the conditional covariance matrix $\bm{V}_+$ as functions of time. 
The following parameters were used in the plot:
as shown in Table \ref{tab:my_label}.
$\Lambda$ is defined as $m=\rho L^3\Lambda$, where 
$\rho$ is the mass density of each mirror.
Fig.~\ref{fig:v} demonstrates that the system converges to the steady-state solution for $t> 700~[{\rm s}]$, which is described by the static solution of the Riccati equation in \cite{miki23}:
\begin{eqnarray}
    &&\bar V_{qq}^{\pm}
    =\frac{\gamma_{\pm}-\gamma_{m}}{\lambda_{\pm}},~~~
    \bar V_{qp}^{\pm}
    =\frac{(\gamma_{\pm}-\gamma_m)^2}{2\lambda_{\pm}\Omega_{\pm}},\notag\\
    &&\bar V_{pp}^{\pm}
    =\frac{(\gamma_{\pm}-\gamma_{m})(2\Omega_{\pm}^2+\gamma_{\pm}^{2}-\gamma_{m}\gamma_{\pm})}{2\lambda_{\pm}\Omega_{\pm}^2}.
    \label{sss}
\end{eqnarray}
Here
$\gamma_{\pm}
    =\sqrt{\gamma_m^2-2\Omega_{\pm}^2+2\Omega_{\pm}\sqrt{\Omega_{\pm}^2+\bar{n}_{\pm}\lambda_{\pm}}}$ and 
$\lambda_{\pm}
={16g_{\pm}^2}/{\kappa}$.
This solution was first obtained by using the Wiener filter described in Ref.~\cite{Bowen20}.
The early phase solution for $t< 10~[{\rm s}]$ is approximately described  by 
$
V_{qq}^\pm(t)=\kappa/\left(16tg_\pm^2\right),\quad
V_{qp}^\pm(t)=\Omega_\pm\left(n_{\rm th}^\pm+1/2\right) t. \quad
V_{pp}^\pm(t)=2n_{\rm th}^\pm+1
$,
where $V_{pp}^\pm(t)$ and $V_{qp}^\pm(t)$ are determined using the initial values of $V_{pp}^\pm(0)$. However, $V_{qq}^\pm(t)$ is an attractor solution that does not depend on the initial conditions, 
$V_{qq}^\pm(0)$. 
Subsequently, the solution for $30[s]<t<300[s]$ can be approximated as
$
V_{qq}^\pm(t)=\kappa/\left(4tg_\pm^2\right),\quad
V_{qp}^\pm(t)=3\kappa/\left(8t^2g_\pm^2\Omega_\pm\right),\quad
V_{pp}^\pm(t)=3\kappa/\left(8t^3g_\pm^2\Omega_\pm^2\right)
$,
where all components are attractor solutions, which 
did not depend on initial conditions.
The components $V_{qq}^+(t)$ and $V_{pp}^+(t)$ are significantly reduced 
from the initial values determined using the thermal phonon number. 
Hence, the mechanical modes become quantum-squeezed states through the Kalman filter process.

\begin{figure}
    \centering
    \includegraphics[width=8.cm]{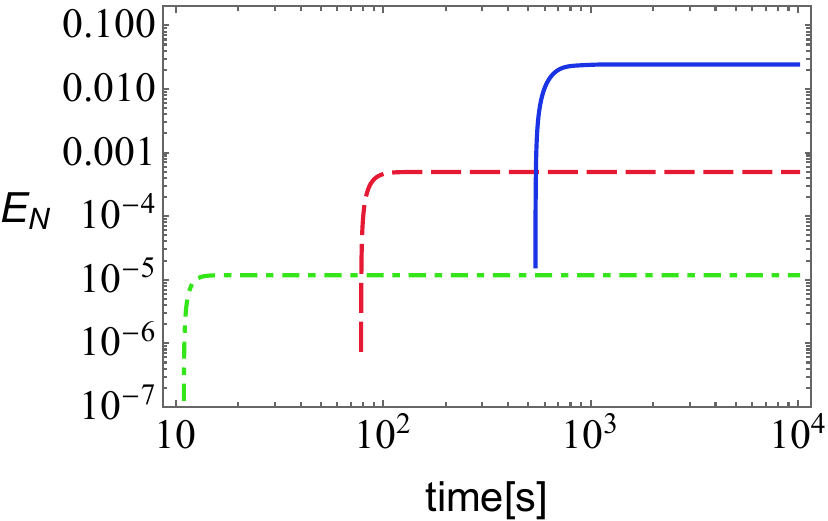}
    \caption{  Behavior of the logarithmic negativity between mechanical mirrors under the Kalman filtering.
    Each curve assumes $\kappa/2\pi=$ $10^8 {\rm Hz}$ (blue solid curve), 
    $10^7 {\rm Hz}$ (red dashed curve), $10^6 {\rm Hz}$ (green dashed-dotted curve), 
respectively. The others are the same as those in Table~\ref{tab:my_label}. 
\label{fig:p}}
\end{figure}

We focused on the GIE between individual mirrors A and B. 
Using the operation of the half-beam splitter, the conditional covariance matrix 
$\bm{\mathcal{V}}$ of the individual mirrors were obtained. 
The entanglement between the mirrors is quantified by the logarithmic negativity 
$E_N$, which is given by the mirror covariance matrix. 
Appendix A presents a method to obtain 
$\bm{\mathcal{V}}$ and 
$E_N$ from 
$\bm{V}_\pm$ via the beam splitter operation.
Fig.~\ref{fig:p} shows the behavior of $E_N$ as a function of time, and we observe the generation of 
$E_N$, that is, the generation of entanglement.    
From the solution (\ref{sss}), we evaluated the logarithmic 
negativity in the steady state, as
\begin{eqnarray}
E_N&\simeq
       \displaystyle{ -\frac{1}{2}\log_2
    \Bigl[
    1-\frac{\kappa\gamma_m}{16g_+^2}\Bigl(\frac{\Omega}{\sqrt{2}\gamma_m}\epsilon-4n_{\rm th}^+\Bigr)\Bigr]},
    \label{anEn}
\end{eqnarray}
where the approximations $\bar{n}_+\lambda_+\gg\Omega_+^2\gg\gamma_m^2$ and $4g_+^2/\kappa\gamma_m\gg n_{\rm th}^+\ge1$ were used.
The entanglement criterion is as follows:
$
    \Omega\epsilon>
    4\sqrt{2}\gamma_mn_{\rm th}^+,
$
which is nearly the same as the entanglement criterion between the output lights \cite{Miao20,Datta21}.
Entanglement appears after the purity of the mirror state increases, and the 
asymmetry of squeezed states between the common and differential modes
is necessary for entanglement generation. Appendix B presents the time evolution of the squeezing properties.

Entanglement does not occur without quantum control or optimal filtering for the initial thermal state with $n_{\rm th}^\pm\simgt10$ 
(see Ref.~\cite{Krisnanda}).
The numerical solutions shown in Fig.~\ref{fig:p} converge to the following analytical formula 
(\ref{anEn}). 
Timescale to achieve an entangled steady state
does not depend on the initial state but changes roughly in proportion to $\kappa/g_+^2$. 
Ref.~\cite{Krisnanda} showed the time scale for generating GIE between two harmonic oscillators without the input light of optomechanical coupling, 
$t_{\rm en}=\pi/(\Omega\epsilon)
\nonumber =1.8\times 10^3 
\Bigl(\frac{\Omega/2\pi}{10^{-3} {\rm Hz}}\Bigr)
    {\Bigl(\frac{\rho}{20{\rm g/cm^3}}\Bigr)}^{-1}
    \Bigl(\frac{\Lambda}{2}\Bigr)^{-1}{\rm s}$. 
This should be compared with the time required to generate an entanglement negativity
in Fig.~\ref{fig:p}, which was shorter than $t_{\rm en}$, particularly for 
small $\kappa$. 
Therefore, the use of the quantum Kalman filter is advantageous
to generate entanglement and preserve the quantum coherence of the system.

Finally, we discuss the error in the measurement of $E_N$, which can be 
estimated by 
\begin{eqnarray}
&&\hspace{-5mm}
\Delta E_N(V)
\nonumber\\
&&\hspace{-3mm}=
\sum_{j=\pm}\Bigl({\partial E_N\over \partial{\bar V^j_{qq}}}\Delta V^j_{qq}
+{\partial E_N\over \partial{\bar V^j_{qp}}}\Delta V^j_{qp}
+{\partial E_N\over \partial{\bar V^j_{pp}}}\Delta V^j_{pp}\Bigr). 
\end{eqnarray}
Assuming that the errors $q_\pm$ and $p_\pm$ 
follow a Gaussian distribution around the estimated values, 
we may write
$\langle\Delta V_{qq}^2\rangle=2\bar V_{qq}^2
$, 
$\langle\Delta V_{qp}^2\rangle=\bar V_{qq}\bar V_{pp}+\bar V_{qp}^2
$, 
$\langle\Delta V_{pp}^2\rangle=2\bar V_{pp}
$, 
$\langle\Delta V_{qq}\Delta V_{qp}\rangle=2\bar V_{qq}\bar V_{qp}
$, 
$\langle\Delta V_{qq}\Delta V_{pp}\rangle=2\bar V_{qp}^2$. 
Then, the variance in the error of $\Delta E_N(V)$ is given by
\begin{eqnarray}
\hspace{-5mm}&&\langle \Delta E_N^2(V)\rangle=
\nonumber\\
&&
2\Bigl({\partial E_N\over \partial {\bar V^+_{qq}}}\Bigr)^2\bar V^{+2}_{qq}
+\Bigl({\partial E_N\over \partial {\bar V^+_{qp}}}\Bigr)^2(\bar V^{+2}_{qp}+\bar V^+_{qq}\bar V^+_{pp})
\nonumber\\
&&+2\Bigl({\partial E_N\over \partial {\bar V^+_{pp}}}\Bigr)^2\bar V^{+2}_{pp}
+4\Bigl({\partial E_N\over \partial {\bar V^+_{qq}}}\Bigr)
\Bigl({\partial E_N\over \partial {\bar V^+_{pp}}}\Bigr)
\bar V^{+2}_{qp}
\nonumber\\
&&+4\Bigl({\partial E_N\over \partial {\bar V^+_{qq}}}\Bigr)
\Bigl({\partial E_N\over \partial {\bar V^+_{qp}}}\Bigr)\bar V^+_{qq}\bar V^+_{qp}
+4
\Bigl({\partial E_N\over \partial {\bar V^+_{pp}}}\Bigr)
\Bigl({\partial E_N\over \partial {\bar V^+_{qp}}}\Bigr)\bar V^+_{pp}\bar V^+_{qp}
\nonumber\\
&&+2\Bigl({\partial E_N\over \partial {\bar V^-_{qq}}}\Bigr)^2\bar V^{-2}_{qq}
+\Bigl({\partial E_N\over \partial {\bar V^-_{qp}}}\Bigr)^2(\bar V^{-2}_{qp}+\bar V^-_{qq}\bar V^-_{pp})
\nonumber\\
&&+2\Bigl({\partial E_N\over \partial {\bar V^-_{pp}}}\Bigr)^2\bar V^{-2}_{pp}
+4\Bigl({\partial E_N\over \partial {\bar V^-_{qq}}}\Bigr)
\Bigl({\partial E_N\over \partial {\bar V^-_{pp}}}\Bigr)\bar V^{-2}_{qp}
\nonumber\\
&&+4\Bigl({\partial E_N\over \partial {\bar V^-_{qq}}}\Bigr)
\Bigl({\partial E_N\over \partial {\bar V^-_{qp}}}\Bigr)\bar V^-_{qq}\bar V^-_{qp}
+4
\Bigl({\partial E_N\over \partial {\bar V^-_{pp}}}\Bigr)
\Bigl({\partial E_N\over \partial {\bar V^-_{qp}}}\Bigr)\bar V^-_{pp}\bar V^-_{qp}.
\nonumber\\
\end{eqnarray}
Assuming that we perform ${\cal N}$ measurements of the GIE, 
we estimated the signal-to-noise ratio as 
$    {\rm S/ N}={\sqrt{\cal N}E_N/ \Delta E_N}$. 
To achieve ${\rm S/N}=1$, the required number of measurements is
$    {\cal N}={\langle\Delta E_N^2(V)\rangle/ E_N^2 }$. 
Provided that the condition $\Omega\epsilon\gg4\sqrt{2}\gamma_mn^+_{\rm th}$ is satisfied, $E_N$ is approximately estimated as $E_N\simeq \kappa G\rho/(8\sqrt{2}g_+^2\Omega)$. 
However, perturbation analysis around the steady state
provides the generation time of the entanglement
with Kalman filtering in proportion to 
$\kappa/g_+^2$.
Subsequently, the total time $\tau$ required to achieve S/N$=1$
is proportional to $\kappa/(g_+E_N)^2 
\propto\omega_cP_{\rm in}\Omega/(m\ell^2\kappa^2)$, where we assume that $\Delta E_N
$ is a constant of ${\cal O}(1)$.
For the parameters in Table I, $\tau\simeq 2\times 10^6$~[s].

We investigated the feasibility of detecting GIE using optomechanical systems. 
For the first time, we determined a feasible set of experimental parameters for achieving a signal-to-noise ratio $S/N$=1.
This is achieved by solving the Riccati equation to determine the time evolution of the conditional state of macroscopic mechanical mirrors in optomechanical systems under quantum control using the Kalman fingering process.
When the two mirrors are coupled via gravitational interaction, the quantum-squeezed states for the common and differential modes are slightly different, which gives rise to the GIE. 
This timescale, associated with
the quantum state squeezing using the Kalman filtering process, is faster than the well-known timescale $t_{\rm en}$.
The parameters in Table I can be an optimal set.  
The feasibility of experiments for detecting the GIE should be investigated in more detail in the future.
The rapid development of GIE is an advantage of optomechanical systems.

{\it acknowledgment }
We thank Nobuyuki Matsumoto, Satoshi Iso, Katsuta Sakai, Kiwamu Izumi for valuable discussions related to the topic of the present paper.
This work was supported by JSPS KAKENHI, Grant No.~JP22J21267 (D.M.),  Nos.~JP23K13103 and JP23H01175 
(A.M.), and No.~JP23H01175 (K.Y.).

\onecolumngrid

\section{Appendix A: Covariance matrix of individual mirrors}

Formula for covariance matrix 
$\bm{\mathcal{V}}$ for the individual mirrors is expressed as follows: 
\begin{eqnarray}
    \bm{\mathcal{V}}&\equiv
    \Bigl(\begin{array}{cc}
\bm{\mathcal{V}}_A&\bm{\mathcal{V}}_{AB}\\
        \bm{\mathcal{V}}_{AB}&\bm{\mathcal{V}}_{B}
    \end{array}\Bigr)
    =\bm{S}
    \Bigl(\begin{array}{cc}
        \bm{V}_+&0
        \label{bmV}
        \\
        0&\bm{V}_{-}
    \end{array}\Bigr)
    \bm{S}^{\text{T}},\qquad
    \bm{S}=\frac{1}{\sqrt{2}}
    \left(\begin{array}{cccc}
            1&0&1/(1-\epsilon)^{1/4}&0\\
        0&1&0&(1-\epsilon)^{1/4}\\
        1&0&-1/(1-\epsilon)^{1/4}&0\\
        0&1&0&-(1-\epsilon)^{1/4}
    \end{array}\right),
\end{eqnarray}
where 
$\epsilon=4Gm/(L^3\Omega^2)$, 
$\bm{S}$ denotes the operation of the half-beam splitter and 
$\bm{V}_\pm$ are the covariance matrices of common and differential modes, respectively. 
The covariance matrices are given as solutions to Riccati equation \eqref{eq:Riccati}.
Here, $\bm{\mathcal{V}}_A$ and $\bm{\mathcal{V}}_B$
are the covariance matrices of mirrors A and B normalized by the frequency $\Omega$, respectively. 
$\bm{\mathcal{V}}_{AB}$ represents the gravity-induced correlation matrix between individual mirrors.
We then introduce the logarithmic negativity $E_N$ to investigate the entanglement as
 $  E_{N}=
    -\frac{1}{2}\log_2\left[{(\Sigma-\sqrt{\Sigma^{2}-4\text{det}\bm{\mathcal{V}}})}/{2}\right]$,
where $\Sigma=\det\bm{\mathcal{V}}_{A}+\det\bm{\mathcal{V}}_{B}-2\det\bm{\mathcal{V}}_{AB}$.
According to the separability condition for two-mode Gaussian states, the systems are entangled if and only if $E_N>0$ \cite{Giedke2001}.

\section{Appendix B: Purity and squeezing}

\begin{figure}[H]
\includegraphics[width=5.6cm]{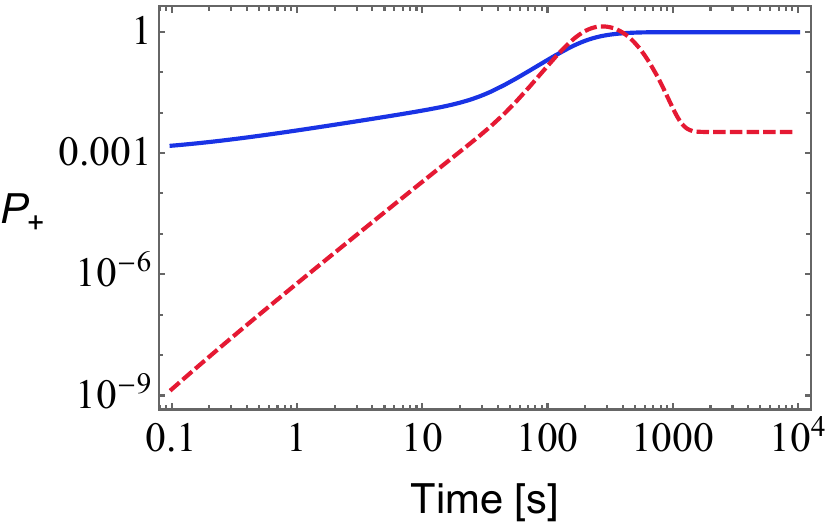}~~~~
    \includegraphics[width=5.6cm]{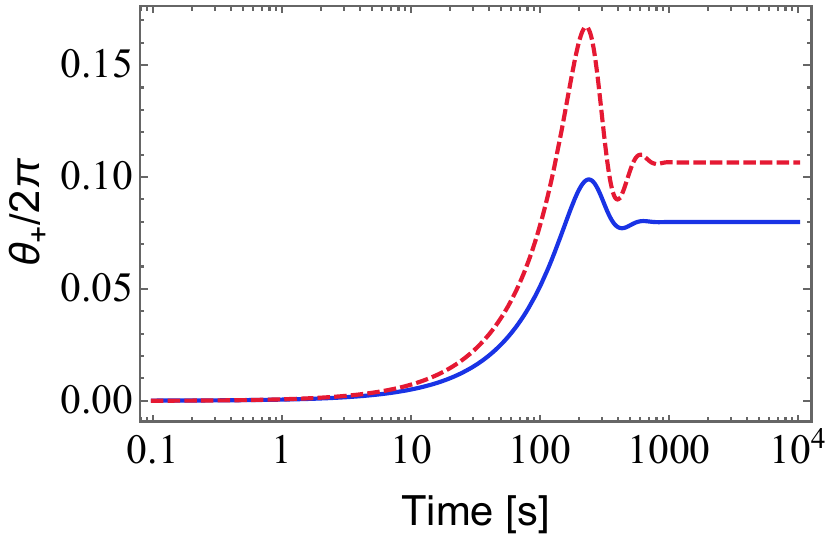}
    ~~~~
    \includegraphics[width=5.4cm]{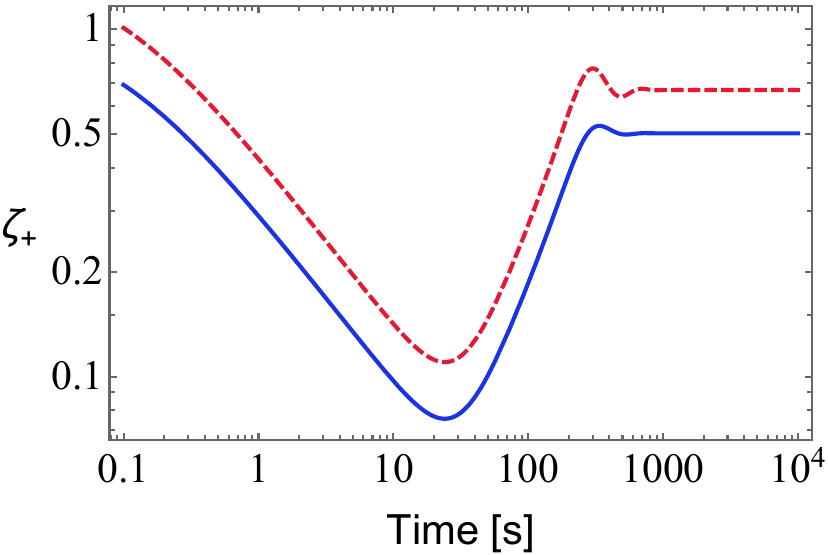}
\caption{ The blue curve in each panel shows the purity of the mechanical common mode (left panel), the squeezing angle of the mechanical common mode (center panel), the ratio of the minimum eigenvalues to maximum eigenvalues of the mechanical covariance matrix $\bm{V}_+$ (right panel).
The red dashed curve in each panel shows the difference between the common mode and the differential mode multiplied by $10^2$ in the left panel and $10$ in the center and right panels.
The parameters used in this figure are the same as those in Table~I.
\label{fig:pc}
}
\end{figure}

Here, we demonstrate the time evolution of the conditional state realized using the Kalman filtering process. The solid blue curves in the panels in Fig. ~\ref{fig:pc} show the evolution of the common mode state, whereas the red dashed curve shows the difference between the common and differential modes.
The left panel of  Fig.~\ref{fig:pc} shows purity as a function of time. 
The center and right panels show 
squeezing angle and degree of squeezing, respectively, as a function of time.

\end{document}